\documentclass[multphys,vecphys]{svmult}

\usepackage{makeidx}         
\usepackage{graphicx}        
\usepackage{multicol}        
\usepackage[bottom]{footmisc}

\makeindex             

\begin{document}

\title*{Integral-field studies of the high-redshift Universe}
\author{Matt J.~Jarvis\inst{1}\and
Caroline van Breukelen\inst{1}\and
Bram P.~Venemans\inst{2}\and
R.~J.~Wilman\inst{3}}
\authorrunning{Jarvis M.J. et al.} 
\institute{Astrophysics, Department of Physics, Keble Road, Oxford, OX 3RH, U.K. \texttt{m.jarvis1@physics.ox.ac.uk, cvb@astro.ox.ac.uk}
\and  
Sterrewacht Leiden, PO Box 9513, 2300 RA Leiden, The Netherlands \texttt{venemans@strw.leidenuniv.nl} \and
Department of Physics, University of Durham, Durham DH1 3LE, U.K. \texttt{r.j.wilman@durham.ac.uk} 
}
%
%
\maketitle

\begin{abstract}
We present results from a new method of exploring the distant
Universe. We use 3-D spectroscopy to sample a large cosmological
volume at a time when the Universe was less than 3 billion years old
to investigate the evolution of star-formation activity.  Within this study we also discovered a high redshift type-II quasar which would not have been identified with
imaging studies alone. This highlights the crucial role that
integral-field spectroscopy may play in surveying the distant Universe
in the future.
\end{abstract}

\section{Hunting for high-redshift galaxies with IFUs}
We initiated a pilot project with a deep, nine hour, VIMOS observation
centred on the high-redshift radio galaxy MRC0943-242 at a redshift of
$z=2.92$ in April 2003. The aims of this project were to probe the
giant-Ly$\alpha$ emitting halo surrounding this source and
the distribution of galaxies within the volume probed by the
IFU. With its spectral coverage and large field-of-view, VIMOS is currently the best IFU for such studies.
We are able to detect all the galaxies with bright
emission lines over the whole volume. For Ly$\alpha$
emission this range is $2.3 < z < 4.6$, and for [OII]$\lambda 3727$ emission, we probe $0.08<z< 0.83$. Therefore we can search for emission-line galaxies over a
large fraction of cosmic volume along the sightline of the IFU.

The process of detecting and selecting emission line objects from IFU data is difficult and involves multiple steps, therefore we refer the reader to van Breukelen, Jarvis \& Venemans (2005) for full details.  However, our selection enabled us to detect 17 emission-line objects over the
volume probed with the IFU. These are predominantly single line
objects, and for 14 all of the characteristics point to them being
hydrogen Ly$\alpha$ emission-line galaxies (two others are [OII]
emitters and the third is the type-II quasar discussed later in this
article), we will now concentrate on these Ly$\alpha$
emitters. Ly$\alpha$ emission is produced by massive stars
photoionizing hydrogen gas. By using some simple assumptions it is
possible to estimate the star-formation rate in galaxies which exhibit
Ly$\alpha$ emission by measuring the luminosity of the emission
line. Although undoubtedly crude, this does at least produce a lower
limit for the star-formation activity in distant galaxies. By binning all of the Ly$\alpha$ luminosities in the volume we construct the Ly$\alpha$ emitter luminosity function. Construction of the luminosity function is a non-trivial task
for this type of data because those galaxies with bright emission line
can be seen to much greater distance in the volume covered in our
data, thus the volume probed is a strong function of the luminosity of
the emission lines and an accurate senisitivity map of the field is crucial (see van Breukelen et al. 2005). 

\begin{figure}
\centering
\includegraphics[height=4cm]{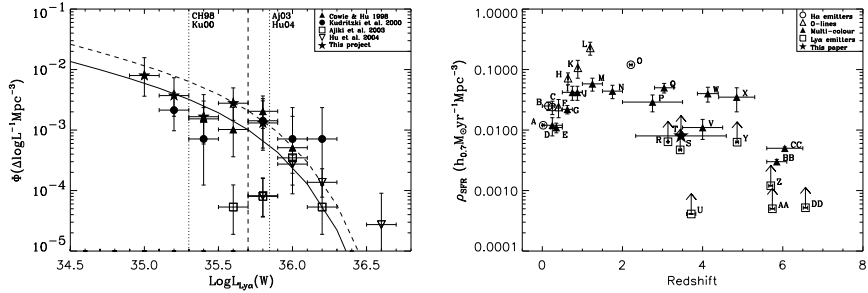}
\caption{({\it left}) The number density of Ly$\alpha$ emitters plotted against the
luminosity. The filled symbols mark surveys with an average redshift
similar to ours (triangles and circles) and the open symbols stand for
surveys at redshift $z=5.7$ (squares and inverted
triangles). Overplotted are two Schechter luminosity functions: the
solid line is the fit to all our data points and the dashed line is
the fit to our two highest luminosity data points and those of the
surveys at similar redshift with $L > 5 \times 10^{35}$~W (dashed horizontal
line) to ensure completeness. The dotted horizontal lines mark the
detection limits of the surveys. ({\it right}) Star-formation rate densities as derived by various types of surveys. The result from our work is denoted by the filled star. The different types of surveys are marked with different symbols: the open circles are H$\alpha$ searches, the open triangles are surveys aimed at oxygen emission lines, the filled triangles are multicolour surveys, and the open squares are Ly$\alpha$ searches (full details of the other surveys can be found in van Breukelen et al. 2005).
}
\label{fig:jarvislumfunc}       
\end{figure}

Fig.~\ref{fig:jarvislumfunc} ({\it left}) shows the Ly$\alpha$ luminosity function derived
from this study compared to the luminosity function measured from
narrow-band studies and multi-colour selection. One can see that our
luminosity function, which probes the redshift range $2.3 < z < 4.6$
extends the work of the narrow-band searches to fainter luminosities
where the luminosity function keeps the same Schechter function form up to $z \sim 6$. This implies that there is little evolution
in the star-formation rate density over this redshift range, albeit
small number statistics preclude strong statements regarding any
evolution.

As stated above, knowledge of the luminosity of the Ly$\alpha$
emission line in these galaxies informs on the total star-formation
rate. Using typical assumptions of hydrogen recombination the
star-formation rate is related to the Ly$\alpha$ luminosity by $SFR = 9.1 \times 10^{-36} (L_{\rm Ly\alpha} / {\rm W})$~M$_{\odot}$~yr$^{-1}$ [see van Breukelen et al. (2005) for details].
By integrating over the Ly$\alpha$ luminosity function we are
therefore able to measure the star-formation rate at the redshifts
covered by our data. This plot, along with the star-formation rate
density derived by other methods, is shown in Fig.~\ref{fig:jarvislumfunc} ({\it right}) for $0 < z <
6$. Due to the fact that Ly$\alpha$ can be resonantly
scattered and absorbed by neutral hydrogen around the source, the
measured SFR from studies using Ly$\alpha$ are hard lower
limits. Also, the presence of dust preferentially extinguishes the UV
continuum emission, therefore even multi-colour searches are prone to
biases which work to reduce the estimated SFR. Therefore, we also show
the estimated star-formation rate corrected for obscuration. With this
correction in place it is apparent that our IFU search is in line with
previous studies conducted in a number of different ways. However, the
benefit of using the integral-field approach is that we select sources
at all redshifts in our volume in precisely the same way, thus
reducing the biases involved in comparing studies at different
redshifts from different surveys, which may utilize different
techniques.

\section{Discovery of a type-2 quasar in the IFU deep field}

\begin{figure}
\centering
\includegraphics[height=5cm]{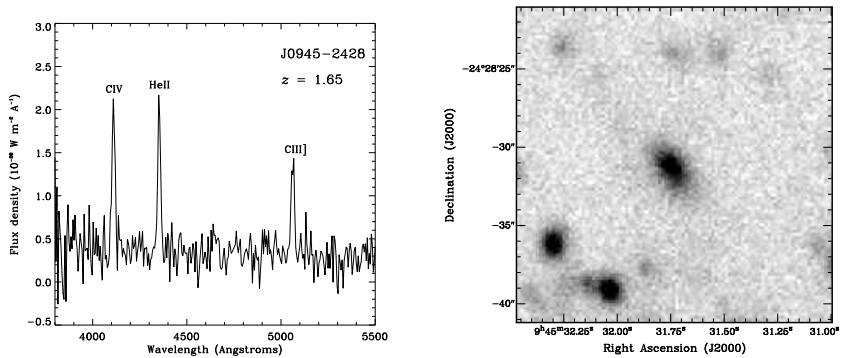}
%
%
\caption{ ({\it left}) The 1-D spectrum of the type-II quasar,
    J0945-2428 at $z = 1.65$. The spectrum was extracted over the
    whole galaxy. ({\it right}) V-band image of J0945-2428, clearly showing that the quasar is resolved.}
\label{fig:jarvistype2}       
\end{figure}

In this section we discuss the way in which our integral-field data has
also led to the discovery of two Active Galactic Nuclei (AGN) in the
volume probed, in addition to the radio galaxy which was targeted. One
of these is a `normal' unobscured type-I quasar with broad emission
lines and an unresolved morphology on optical images at a redshift of
$z=1.79$. However, the other AGN exhibits only narrow-emission lines
[Fig.~\ref{fig:jarvistype2} ({\it left})] and has a resolved morphology in the optical image [Fig.~\ref{fig:jarvistype2} ({\it right})].

These type-II AGNs are relatively difficult to find compared to the
type-I counterparts. This is principally due to the fact that type-II
AGN look like normal galaxies, and it is only by looking for other
signatures of AGN activity, which do not suffer from extinction due to
the torus, can they be found, e.g. X-rays from the central engine
which penetrate the torus (e.g. Norman et al. 2002), radio emission from powerful jets (e.g. Jarvis et al. 2001a) or
reprocessed dust emission in the mid-infrared from the torus
itself (e.g. Mart\'\i nez-Sansigre et al. 2005). However, with the integral-field approach we are sensitive to
the bright narrow-emission lines that are characteristic of an
obscured AGN, as we obtain the spectrum of any object in the IFU field
immediately.

J094531-242831 (hereafter J0945-242) exhibits these bright
narrow-emission lines, in the CIV doublet, HeII and CIII], all
characteristic of a type-II AGN. The radio map shows that there is no
radio emission down to a radio flux limit of 0.15 mJy at 5GHz. At a
redshift of $z=1.65$ this is significantly below the typical luminosity
of a radio galaxy, thus we confirm that this is a genuine radio-quiet
type-II quasar. The line luminosity ratios of the CIV, HeII and CIII]
lines are also consistent with the ratios for radio galaxies, and not
the generally lower-luminosity Seyfert-I galaxies and the unobscured
quasars (McCarthy 1993). Using these line luminosities it is possible to estimate the
lower mass limit of the accreting black hole in the centre of this
galaxy. We assume the typical line ratios of radio galaxies to convert
the Hell luminosity to a line luminosity in [OII], which is correlated
with the total bolometric luminosity of the AGN. Under the assumption
that the quasar is accreting at its maximum rate, i.e. the Eddington
limit, then this bolometric luminosity equates to a black-hole mass of
$3 \times  10^{8}$~M$_{\odot}$.

In the local Universe there is now a well known correlation between
the mass of black holes and the luminosity of their host galaxy (see
e.g. Magorrian et al. 1998). The near-infrared $K-$band magnitude of
J0945-242 is very faint, with $K=20.5$. Radio galaxies at $z=1.65$
typically have host galaxy luminosities of $K \sim 18$ (e.g. Jarvis et al. 2001b). Thus the host galaxy
of J0945-242 appears to be $2.5$~mag fainter than
that for a typical radio-loud type-II AGN. If this faintness of the
host galaxy is caused by extinction from dust then we would expect the
blue end of the galaxy spectrum to be fainter, as dust attenuates the
blue light more readily than at red wavelengths. However, the host
galaxy of J0945-242 is extremely blue, indicative of ongoing
star-formation. Therefore, the faintness in the $K-$band light indicates
that the host galaxy has a dearth of old, massive stars, which in turn
implies that the galaxy is not yet fully formed at $z=1.65$. Whereas the
black hole has already grown, presumably by accretion of matter, close
to its final mass due to the fact that the low-redshift black-hole
mass function shows that supermassive black holes appear to have a
maximum mass of around $10^{10}$~M$_{\odot}$ (e.g. Marconi et al. 2004).

This relatively large black-hole mass associated with a host galaxy
approximately a factor of 10 fainter than would be expected from
the local relation implies that supermassive black holes at high
redshift may essentially be fully grown before the host galaxy has
fully formed. This is in qualitative agreement with what we already
see in high-redshift radio galaxies, where the small, young, radio
sources appear to have extremely bright sub-millimetre
luminosities (e.g. Archibald et al. 2001).
In order to produce these
sub-millimetre luminosities, star-formation rates of up to 1000
M$_{\odot}$~yr$^{-1}$ are needed, typical of a galaxy undergoing its
first major bout of star formation activity.

\section{Summary}
The new method of detecting emission-line galaxies at high
redshift along with the serendipitous discovery of an obscured quasar
at $z=1.65$, highlights the way in which relatively wide-area
integral-field units on large telescopes could open up a unique window
on the Universe. VIMOS is currently the only instrument which has the
capability of large spectral coverage coupled with a ~1 square
arcminute field-of-view. However, future instruments, such as the
Multi-Unit Spectroscopic Explorer (MUSE; http://muse.univ-lyon1.fr),
will expand the initial work taking place in this field with
VIMOS. Furthermore, volumetric surveys with IFUs may begin to find
types of object we have yet to discover in traditional surveys, and
thus offer a whole new view of the Universe.

Full details of the work presented in this article can be found in van Breukelen et al. (2005) and Jarvis, van Breukelen \& Wilman (2005).



\printindex
\end{document}